\RequirePackage{tikz}
\RequirePackage{pgf}

\documentclass[pdflatex,sn-mathphys]{sn-jnl}

\usepackage{psfrag}
\usepackage{mathrsfs}
\usepackage{amssymb, bm}
\usepackage{amsmath, amsthm}
\usepackage{epstopdf}
\usepackage{hyperref}
\usepackage{enumerate}
\usepackage{longtable}
\usepackage{subfigure}
\usepackage{color}
\usepackage{mathrsfs}
\usepackage{graphicx}
\usepackage{bm,natbib,url,textcase}
\usepackage[english]{babel}
\usepackage[T1]{fontenc}
\usepackage{comment}

\usepackage{multirow}
\usepackage{stackengine}
\newcommand\xrowht[2][0]{\addstackgap[.5\dimexpr#2\relax]{\vphantom{#1}}}
\usepackage[normalem]{ulem}
\useunder{\uline}{\ul}{}
\usepackage{booktabs}
\usepackage{hhline}
\usepackage{tabularx}
\usepackage{nicematrix}
\usepackage{color}

\def\av{``} 
\def\cv{'' } 
\def\z{~} 
\def\nn{\nonumber} 
\def\T{{\cal T}}
\def\K{{\cal K}}

\def\eps{\varepsilon}

\theoremstyle{thmstyleone}%
\theoremstyle{thmstyletwo}%

\theoremstyle{thmstylethree}%

\begin{document}

\title[Fluid nature constrains Horndeski gravity]{Fluid nature constrains Horndeski gravity}


\author*[1,2]{\fnm{Marcello} \sur{Miranda}}\email{marcello.miranda@unina.it}


\author[3,2,1]{\fnm{Daniele} \sur{Vernieri}}\email{daniele.vernieri@unina.it}
\equalcont{These authors contributed equally to this work.}


\author[3,2,1]{\fnm{Salvatore} \sur{Capozziello}}\email{capozziello@na.infn.it}
\equalcont{These authors contributed equally to this work.}


\author[4]{\fnm{Valerio} \sur{Faraoni}}\email{vfaraoni@ubishops.ca}
\equalcont{These authors contributed equally to this work.}


\affil*[1]{\orgname{Scuola Superiore Meridionale}, \orgaddress{\street{ Largo San Marcellino 10}, \city{Napoli}, \postcode{I-80138}, \state{Italy}}}

\affil[2]{\orgname{Istituto Nazionale di Fisica Nucleare, Sez. di Napoli}, \orgaddress{\street{Compl. Univ. Monte S. Angelo, 
Edificio G, Via	Cinthia}, \city{Napoli}, \postcode{I-80126}, \state{Italy}}}

\affil[3]{\orgdiv{Dipartimento di Fisica ``E. Pancini''}, \orgname{Universit\`{a} di 
Napoli ``Federico II''}, \orgaddress{\street{Compl. Univ. Monte S. Angelo, 
Edificio G, Via	Cinthia}, \city{Napoli}, \postcode{I-80126}, \state{Italy}}}

\affil[4]{\orgdiv{Department of Physics \& Astronomy}, \orgname{Bishop's University}, \orgaddress{\street{2600 College Street}, \city{Sherbrooke}, \postcode{J1M~1Z7}, \state{Qu\'ebec}, \country{Canada}}}


\abstract{The elusive physical nature of Horndeski gravity is elucidated in a new approach depicting this class of theories as a dissipative effective fluid. Requiring the constitutive equations of the latter to be those of a Newtonian fluid restricts the theory to only two disconnected subclasses of ``viable'' Horndeski gravity. Therefore, a stress-energy tensor of the Horndeski effective fluid, linear in the first derivatives of the fluid's 4-velocity, is a sufficient condition for gravitational waves to propagate at light speed. All other Horndeski theories correspond to exotic non-Newtonian effective fluids. The two linear Horndeski classes are studied in the framework of first-order thermodynamics of viscous fluids, which further constrains the functional form of the theory.}




\maketitle

\section{Introduction}\label{sec:1}

Einstein's General Relativity (GR) cannot be the ultimate theory of gravity for several reasons. Any 
attempt to reconcile it with quantum mechanics introduces deviations from GR. 
Moreover, the need to explain the current acceleration of the universe without a 
completely {\em ad hoc} dark energy has led to modifications of GR on cosmological 
scales. The most popular class of theories for this purpose is probably $f(R)$ 
gravity\z\cite{Capozziello:2002rd}, a subclass of scalar-tensor gravity\z\cite{Faraoni:2010pgm}. In the last 
decade, 
scalar-tensor gravity has been generalized by rediscovering Horndeski theory
\cite{Horndeski:1974wa} (see Ref.\z\cite{Kobayashi:2019hrl} for a recent review), leading to a 
flurry of activity. Horndeski gravity is one of the most general scalar-tensor theories with second-order field equations, which avoids the notorious Ostrogradsky instability 
\cite{Ostrogradsky:1850fid,Woodard:2015zca}. The action is commonly written as\z\cite{Kobayashi:2019hrl}
\begin{equation}
S\left[ g_{ab}, \phi \right] = \int d^4 x \sqrt{-g} \, \left( {\cal L}_2 + 
{\cal L}_3+ {\cal L}_4+ 
{\cal L}_5 \right) + S^\mathrm{(m)} \,, \label{Horndeskiaction}
\end{equation}
where $g_{ab}$ is the spacetime metric with determinant $g$, $\phi$
is the scalar degree of freedom, $S^\mathrm{(m)}$ is the matter action,\footnote{Following the notation of 
Ref.~\cite{Wald:1984rg}, we use units in which the speed of light $c=1$ and $8\pi G=1$, where $G$ is  
Newton's constant, the metric signature is ${-}{+}{+}{+}$, and $R$ denotes 
the Ricci scalar.} 
\begin{align}
{\cal L}_2 =\,& G_2\left( \phi, X \right) \,,\qquad 
{\cal L}_3 =\,- G_3\left( \phi, X \right) \Box \phi \,,\nn\\ 
&\nn\\
{\cal L}_4 =\,& G_4\left( \phi, X \right) R +G_{4X} \left( \phi, X \right) 
\left[ \left( \Box \phi \right)^2 -\left( \nabla_a \nabla_b \phi \right)^2 
\right] \,, \nonumber\\ 
&\nn\\
{\cal L}_5  =\,&  G_5\left( \phi, X \right) G_{ab}\nabla^a \nabla^b \phi 
\nonumber\\
&\nn\\
& -\frac{ G_{5X} }{6} \left[ \left( \Box\phi \right)^3 -3\Box\phi 
\left( \nabla_a\nabla_b \phi \right)^2 +2\left( \nabla_a \nabla_b \phi 
\right)^3 \right] \,.
\end{align}
$ X \equiv -\tfrac{1}{2}\,\nabla^c\phi \nabla_c\phi  $, the $G_i$   
($i=2,3,4,5$) are regular functions of $\phi$ and $X$, while $ G_{i\phi} 
\equiv \partial G_i/\partial \phi$ and $ G_{iX} \equiv \partial 
G_i/\partial X$, $ \left( \nabla_a\nabla_b \phi \right)^2 
\equiv \nabla_a\nabla_b\phi\nabla^a \nabla^b \phi$ and $ \left( 
\nabla_a\nabla_b \phi \right)^3 \equiv \nabla_a\nabla_c\phi 
\nabla^c\nabla^d \phi \nabla_d \nabla^a \phi $. 

The multi-messenger event GW170817/GRB170817A from a neutron star binary merger 
\cite{LIGOScientific:2017vwq,LIGOScientific:2017zic} restricts Horndeski gravity to 
the so-called ``viable'' class characterized by $
G_5= G_{4X}=0 $, 
in which the speed of gravitational waves equals $c$. This class also avoids 
instabilities and admits an Einstein frame description 
\cite{Creminelli:2017sry,Baker:2017hug,Bettoni:2016mij,Andreou:2019ikc,Kobayashi:2019hrl}.

The field equations of viable Horndeski gravity are
\begin{eqnarray}
&& G_4  \, G_{ab} -\nabla_{a}\nabla_{b}G_4  
+ \left[ \Box G_4 -\dfrac{G_2  }{2} 
-\dfrac{1}{2} \, \nabla_{c} 
\phi\nabla^{c}G_{3} 
\right] g_{ab} \nonumber\\
&& + \frac{1}{2}\left[ G_{3X} \, 
\Box\phi -G_{2X} \right] 
\nabla_{a}\phi\nabla_{b}\phi 
+ \nabla_{(a}\phi \nabla_{b)}G_{3} =T^\mathrm{(m)}_{ab} \,,\label{hfeq}\\
&&\nn\\
&& G_{4,\phi}  R + G_{2,\phi} +G_{2X}  
\Box\phi+\nabla_{c}\phi\nabla^{c}G_{2X} \vspace{7pt}\nn\\
&&\nn\\
&&-G_{3X} (\Box\phi)^2-\nabla_{c}\phi\nabla^c 
G_{3X}  \Box\phi-G_{3X} \nabla^{c} 
\phi\Box\nabla_c\phi\vspace{7pt}\nn\\\
&&\nn\\
 &&+G_{3X}  R_{ab}\nabla^{a} 
\phi\nabla^{b}\phi-\Box G_{3} -G_{3,\phi} \Box\phi=0 \label{heom}\,,
\end{eqnarray}
where $G_{ab}$ is the Einstein tensor and
$T^\mathrm{(m)}_{ab}\equiv-\tfrac{1}{\sqrt{-g}}\,\tfrac{\delta S^\mathrm{(m)}}{\delta g^{ab}}$.
Eq.\z\eqref{hfeq} can always be written as the effective Einstein equation
$G_{ab}=T_{ab}+T^\mathrm{(m)}_{ab}/G_{4} \,,$
where $T_{ab}=T^{(2)}_{ab}+T^{(3)}_{ab}+T^{(4)}_{ab}$ is the effective stress-energy
tensor containing all the deviations from GR,
\begin{align}
    T^{(2)}_{ab}=\,&\frac{1}{2G_4} \left( G_{2X}\nabla_{a}\phi\nabla_{b}\phi+G_{2}\,g_{ab} \right)\,,\\
    &\nn\\
    T^{(3)}_{ab}=&\frac{1}{2G_4} \left( G_{3X} \nabla_{c} X  \nabla^{c} \phi  - 2 X G_{3\phi} \right) g_{ab} \nn\\
    \,& - \frac{1}{2G_4}\left( 2 G_{3\phi} + G_{3X} \Box \phi \right) \nabla _a  \phi \nabla _b \phi- \frac{G_{3X}}{G_4} \nabla_{(a} X \nabla_{b)} \phi\,,\\
    &\nn\\
    T^{(4)}_{ab}=\,&\frac{G_{4\phi}}{G_4}(\nabla_{a}\nabla_{b}\phi-g_{ab}\Box\phi)+\frac{G_{4\phi\phi}}{G_4}(\nabla_{a}\phi\nabla_{b}\phi+2X\,g_{ab})\,.
\end{align}

For clarity of 
illustration, we 
temporarily restrict ourselves to this subclass of Horndeski gravity {\em in vacuo}, but
will later extend our results to the most general Horndeski theory with matter.

Because of the many free functions and terms appearing in the Horndeski action, it is difficult to 
grasp the physical meaning of Horndeski gravity and many works remain formal. One would like to understand better Horndeski gravity from the physical point of view. How can one understand, and classify, the physical deviations from Einstein gravity appearing in these theories?  
Here we propose a new 
approach to this class of theories: by regarding its field equations as effective 
Einstein equations, the gravitational terms other than the Einstein tensor, when 
moved to the right-hand side, assume the form of a {\em dissipative} effective fluid 
\cite{Nucamendi:2019uen} 
(a result familiar in less general scalar-tensor theories\z\cite{Faraoni:2018qdr}). 
This effective fluid approach provides a way to classify Horndeski gravity based on the 
nature of this fluid: the requirement that it is a Newtonian fluid ({\em 
i.e.}, with the viscous stresses depending only on the first derivatives of the fluid's 4-velocity) restricts the scope to two subclasses of viable Horndeski gravity, while more general theories correspond to exotic non-Newtonian effective fluids. Here ``(non-)Newtonian'' refers to standard fluid-dynamical terminology:
we always consider relativistic (effective) fluids.  This classification grasps one of the most basic characteristics of a fluid in the usual, non-relativistic and three-dimensional fluid mechanics. Ordinary fluid behaviour is Newtonian, while more exotic (although still common in nature) non-Newtonian fluids are definitely more complicated. In the absence of other physical ways to classify the nature of Horndeski theories (apart from the well-known distinction between theories in which gravitational waves propagate at light speed and those in which they do not), the behaviour of the effective equivalent fluid serves this purpose. Alternative characterizations of Horndeski theories of gravity from the physical point of view are not contemplated in the (now vast) relevant literature.

We show below that the general Horndeski theory contains only two (``linear
Horndeski'') classes with an effective fluid stress-energy tensor linear in
the four-velocity gradient. They are subclasses of the viable Horndeski class
identified by $G_3=G_{4\phi}\,\ln{X}$ and $G_3=0$, respectively. This result is
applied in the context of Eckart's (or first-order) thermodynamics of relativistic viscous fluids. Out-of-equilibrium contributions to the effective $T_{ab}$
are linear in the gradients of the temperature, chemical potential, and four-velocity. It is possible to constrain the functional form of
$G_2$, which is directly related to the equilibrium pressure of the effective
fluid.

The idea behind the thermodynamic analogy with a first-order viscous fluid loosely originates in Jacobson's idea of modified gravity as a non-equilibrium state in a \av thermodynamics of gravitational theories\cv based on a thermal derivation of the Einstein equations\z\cite{Jacobson:1995ab,Eling:2006aw}, in which classical gravity appears as an emergent phenomenon instead of being fundamental. Thus, GR is associated with an equilibrium state of gravity while any dynamic modified theory (in this case, Horndeski's) is interpreted
as an excited, or non-equilibrium, state.\footnote{However, apart from its spirit, the first-order thermodynamics of scalar-tensor gravity is completely different from Jacobson's thermodynamics of spacetime.} 
The alternative gravity-viscous fluid analogy was developed in previous works \cite{Chirco:2010sw,Faraoni:2021lfc,Giusti:2021sku,Giardino:2022sdv}. Here we derive the heat current density, the \av temperature of
modified gravity\cv associated with the effective fluid ({\em i.e.}, the effective
temperature of the excited state), and its viscosity coefficients by studying a first-order {\em effective} viscous fluid with vanishing
chemical potential in the Eckart frame. While the first linear Horndeski
class is characterized by a unique expression of the temperature, the
temperature associated with the second linear Horndeski class depends on the
function $G_2\left(\phi, X \right)$, or, alternatively, on how the shear viscosity depends on the effective equilibrium pressure. Tables\z\ref{tab:1} and\z\ref{tab:2} summarize our results.

\section{Horndeski effective fluids}
\label{sec:2}

The stress-energy tensor of an imperfect fluid has the well-known form  
\begin{align}
  T^{ab}&=\rho u^{a}u^{b}+q^{a}u^{b} + q^{b} u^{a} +\Pi^{ab}\nn\\
  &=\rho u^{a}u^{b}+Ph^{ab}+q^{a}u^{b} + q^{b} u^{a} +\pi^{ab}\,,
\end{align}
where $u^{a}$ is the fluid 4-velocity ($u_{a}u^{a}=-1$), 
$h_{ab} \equiv g_{ab}+u_{a} u_{b}$ (${h_a}^b$ is the projector onto the 
3-space orthogonal to $u^c$), $\rho = T^{ab}u_{a}u_{b}$ is the energy density, 
$P=\tfrac{1}{3}\,T^{ab}h_{ab}$ is the isotropic pressure, $q^{a}=-T^{cd}u_{c}h_{d}{}^{a}$ 
is the heat flux density, $\Pi^{ab}= h^{a}{}_{c}h^{b}{}_{d}T^{cd}=Ph^{ab}+\pi^{ab}$ is 
the stress tensor, and its traceless part $\pi^{ab}$ describes the anisotropic stresses.
Viscous pressure and anisotropic stresses are assumed to obey  
constitutive laws relating them with the expansion scalar $\Theta \equiv 
\nabla_c u^c$ and the traceless shear tensor $\sigma^{ab} \equiv \tfrac{1}{2} \left( h^{ac}\nabla_{c}u^{b} 
+h^{bc}\nabla_{c}u^{a} \right) -\tfrac{1}{3}\Theta \, h^{ab}$.
In particular, using the decomposition $\nabla_{a}=h_{a}{}^{c}\nabla_{c}-u_{a}u^{c}\nabla_{c}$, one has  $\nabla_{b}u_{a}=\sigma_{ab}+\frac{\Theta}{3} \,h_{ab}+\omega_{ab}-\dot{u}_{a}u_{b}$, where $\omega_{ab}\equiv\nabla_{[b}u_{a]}$ is the vorticity tensor, and $\dot{u}^a \equiv u^c \nabla_c u^a$ is the 4-acceleration of the fluid.

For both non-relativistic and relativistic Newtonian 
fluids, these 
constitutive relations are {\em linear in the 4-velocity gradient} 
\cite{Eckart:1940te}, 
\begin{align}\label{constitutive}
 \pi^{ab}=-2\eta \, \sigma^{ab}\,, \quad \quad P=\bar{P}-\zeta \, \Theta\,,
\end{align}
where $\bar{P}$ is the inviscid pressure while $\eta$ and $\zeta$ are the 
shear and bulk viscosity coefficients, respectively. 
Therefore, the imperfect fluid stress-energy tensor has the form
\begin{align}\label{stress}
  T^{ab}&=\rho u^{a}u^{b}+ \left( \bar{P}- \zeta\Theta \right) h^{ab} + 
q^{a}u^{b}+ q^{b} u^{a} -2\eta\sigma^{ab}\,.
\end{align}
In general, the request of linearity in the first derivatives of the fluid four-velocity endows also the energy density with a viscous contribution. Thus, one could contemplate the additional ``constitutive relation''
\begin{eqnarray}
\rho &=& \bar{\rho}-\xi \, \Theta\,, \label{const3}
\end{eqnarray}
where $\bar{\rho}$ is the inviscid density and $\xi$ is a new viscosity (transport) coefficient. The physical interpretation of this new term should be sought as a resistance to compression and expansion. In principle, this extra viscosity coefficient could be related to the bulk viscosity coefficient.

Our goal consists of characterizing  Horndeski theories based on the physical nature 
(Newtonian or non-Newtonian) of their effective fluid equivalent and on their thermodynamics. 

The Horndeski effective fluid 4-velocity is\z\cite{Giusti:2021sku}
\begin{equation}
u^{a} \equiv\frac{\nabla^{a}\phi}{\sqrt{2X}}\,,
\end{equation}
where we assume that the scalar field gradient is timelike, 
$\nabla_{a}\phi\nabla^{a}\phi<0$. This  identification allows us to rewrite the derivatives of $\phi$ and $X$ in terms of the kinematic quantities associated with the effective fluid: 
\begin{align}
    \nabla_{a}\phi=&\,\sqrt{2X}\,u_{a}\,,\qquad    \nabla_{a}X=\,-\dot{X}\,u_{a}-2X\,\dot{u}_{a}\,,\\
    &\nn\\
    \nabla_{a}\nabla_{b}\phi=&\,\sqrt{2X } \, \nabla_{a}u _{b}-\frac{\dot{X} }{\sqrt{2X }} \, u _{a}u _{b}-\sqrt{2X}\dot{u}_{a}u_{b}\,,\label{ddf}
\end{align}
where $\dot{X} \equiv u^c \nabla_c X$. Then, the effective stress-energy tensor 
associated with viable Horndeski gravity is 
\begin{align}
    T_{ab} =\, & \Bigg[ \frac{2XG_{2X}-G_{2} -2XG_{3\phi}}{2G_4}+\frac{\sqrt{2X}  \left( G_{4\phi} -  X G_{3X} \right)}{  
    G_{4}}\,\Theta \Bigg]u_{a} u_{b}\nn\\
    &+ \Bigg[\frac{ G_{2} + 4 X G_{4\phi\phi}  -2 X G_{3\phi}}{2  
    G_4 } -  \frac{ \left( G_{4\phi} - X G_{3X} 
    \right)}{\sqrt{2X}\,  G_{4} 
    }\,\dot{X} \,-\frac{2 \sqrt{2X} \,  G_{4\phi} \Theta}{3  G_4 }\, \Bigg] h_{ab}\nn\\
    &\nn\\
    &-\frac{2\sqrt{2X} \left( G_{4\phi} - X G_{3X} \right)}{  G_{4}}\dot{u}_{(a} u_{b)}+\frac{\sqrt{2X} \, G_{4\phi} }{ G_{4}} \, \sigma _{ab}\,,
\end{align}
and the associated effective fluid quantities are
\begin{align}
  \rho=\, & \frac{1}{2G_4}\left( 2XG_{2X}-G_{2} -2XG_{3\phi} \right)+\frac{\sqrt{2X}}{G_4} \left( G_{4\phi}-XG_{3X} \right)\Theta\,,\label{rho}\\
  &\nn\\
  P=\,&\frac{1}{2G_4} \left( G_2-2XG_{3\phi}+4XG_{4\phi\phi} \right)-\frac{\left(G_{4\phi}-XG_{3X}\right)}{G_4\sqrt{2X}}\dot{X}-\frac{2G_{4\phi}}{3G_4}\sqrt{2X}\,\Theta\,,\label{P}\\
  &\nn\\
  \eta=\,& -\frac{G_{4\phi}}{2G_4}\, \sqrt{2X}\,, \qquad q^{a} = -\frac{\sqrt{2X}}{G_4} \, \left( G_{4\phi}-XG_{3X} 
\right) \, \dot{u}^{a}\label{q}\,.
\end{align}
One immediately notices that the stress-energy tensor is characterized by anisotropic stresses proportional to the shear tensor $\sigma_{ab}$ and by heat flux density proportional to the fluid's four-acceleration. From the above expressions we cannot obtain all the constitutive relations of the effective fluid associated with viable Horndeski. In Eq.\z\eqref{P}, $\dot{X}$ is not a kinematic quantity. However, the scalar field equation of motion allows us to express it in terms of kinematic quantities.

There are only two possibilities to satisfy the requirement that the total pressure depends {\em linearly} on $\nabla_b u_a$ ({\em i.e.}, that the effective fluid is Newtonian):
\begin{enumerate}
\item The total pressure does not depend on $\dot{X}$, equivalent to
  \begin{equation}\label{case1}
    G_{4\phi} - X G_{3X}=0\quad\Rightarrow\quad 
G_{3}=G_{4\phi}\ln{X}\,.
  \end{equation}

\item The scalar $\dot{X}$ depends only on the scalar field and its kinetic term 
and is linear in the expansion scalar,
  \begin{equation}\label{Xdot}
    \dot{X}=F_1 \left( \phi,X \right) + F_2 \left( \phi,X \right) \, \Theta\,,
  \end{equation}
 where the functions $F_1$ and $F_2$ contribute to the isotropic perfect pressure and the viscous pressure, respectively. $F_{1,2}$ can be determined using the 
scalar field equation of motion. Equivalently, one can work with $\Box\phi$ 
and assume 
 $\Box\phi=\tilde{F}_1(\phi,X)+\tilde{F}_2(\phi,X) \, \Theta$. In either case, 
 $\Box\phi$ or $\dot{X}$ is a function of $\left( \phi , X\right) $ linear
in $\Theta$. We work with $\dot{X}$ without loss of generality.
\end{enumerate}

While the first case directly prescribes the form of $G_3$, in the second
case one has to write $\dot{X}$ in terms of the kinematic quantities to
enforce the linearity in $\nabla_{a}u_{b}$. To do
this, we use  the field equations
\begin{align}\label{feq}
  R_{ab}=\, T_{ab}-\frac{T}{2} \, g_{ab}\,,\qquad R=-\,T\,,
\end{align}
where $T=g_{ab}T^{ab}$ is the trace of the effective stress-energy tensor. The scalar field equation of motion written in terms of the kinematics quantities is 
\begin{align}\label{eom}
    &- \frac{4}{3} X \Theta ^2 G_{3X} + 4 X \sigma ^{2}{} G_{3X} -4 
\dot{u}^{2}{} X \left( G_{3X} + X G_{3XX}\right) -2 \dot{X} \Theta  
\left( G_{3 X} + X G_{3 XX} \right) \nn\\
&+ \frac{\Theta\sqrt{2X}}{G_{4}} \left[ 3 G_{4\phi}^2  -2 X G_{4\phi} 
G_{3X}  -  X^2 G_{3X}^2 + G_{4} \left( G_{2X} -2 G_{3\phi} + 2 X G_{3\phi 
X}\right) \right] \nn\\
& +  \frac{\dot{X}}{\sqrt{2X} G_4} \left\{ 3 G_{4 \phi}^2 -6 X G_{4 \phi} 
G_{3 X} + 3 X^2 G_{3 
X}^2\right.\nn\\
&\left.+ G_4 \left[G_{2 X} + 2 X G_{2 XX} -2 \left(G_{3 \phi} + X G_{3 \phi 
X} \right) \right] \right\} \nn\\
& + \frac{1}{G_4}\left[G_2 \left( -2 G_{4 \phi} + X G_{3 X} \right) + G_4 G_{2 \phi} 
+X^2 G_{3  X} \left( 6 G_{4 \phi\phi} + G_{2 X} -4 G_{3 \phi} 
\right)\right]\nn\\
& +\frac{ X}{G_4} \left[  G_{4 \phi} \left(-6 G_{4 \phi\phi} + G_{2 X} + 2 
G_{3 \phi}\right)  + 2 G_4 \left( - G_{2 \phi X} + G_{3 \phi\phi} \right) 
\right]=0\,,
\end{align}
which can be seen as a linear equation for $\dot{X}$ admitting the algebraic solution
\begin{align}\label{Xdot1}
  \dot{X}=\dfrac{A(\phi,X)+B(\phi,X)\,\Theta+C(\phi,X)\,\Theta^2+D(\phi,X)\,\sigma^2+E(\phi,X)\,\dot{u}^2}{H(\phi,X)+I(\phi,X)\,\Theta}\,,
\end{align}
where
\begin{align}
  A(\phi,X)=\,&\sqrt{2X}\left[G_2 \left( 2 G_{4 \phi} - X G_{3 X} \right) -G_4 G_{2 \phi}\right.\nn\\
  &\left.-X^2 G_{3X} \left( 6 G_{4 \phi\phi} + G_{2 X} -4 G_{3 \phi} \right)\right]\nn\\
  &- \sqrt{2X}\,X \left[ G_{4 \phi} \left(-6 G_{4 \phi\phi} + G_{2 X} + 2 G_{3 \phi}\right)\right.\nn\\
  &\left.+ 2 G_4 \left( - G_{2 \phi X} + G_{3 \phi\phi} \right) \right]\,,\label{AA}\\
  &\nn\\
  B(\phi,X)=\,&2X \left[ 3 G_{4\phi}^2 -2 X G_{4\phi} G_{3X} - X^2 G_{3X}^2\right.\nn\\
  &\left.+ G_{4} \left( G_{2X} -2 G_{3\phi} + 2 X G_{3\phi X}\right) \right] \,,\label{BB}\\
  &\nn\\
  C(\phi,X)=\,&\frac{4}{3} \sqrt{2X}\,X G_{4} G_{3X}\,,\label{CC}\\
  &\nn\\
  D(\phi,X)=\,&-4 \sqrt{2X}\,X G_{4} G_{3X}\,,\label{DD}\\
  &\nn\\
  E(\phi,X)=\,&4 \sqrt{2X}\,X G_{4} (G_{3X} + X G_{3XX})\,,\label{EE}\\
  &\nn\\
  H(\phi,X)=\,& 3G_{4\phi}^2 -6 X G_{4\phi} G_{3X} + 3 X^2 G_{3X}^2+ G_{4} [G_{2X} + 2 X G_{2XX} \nn\\
  &-2 (G_{3\phi} + X G_{3\phi X})]\,,\label{HH}\\
  &\nn\\
  I(\phi,X)=\,&-2 \sqrt{2X} G_{4} (G_{3X} + X G_{3XX})\,,\label{II}
\end{align}
and with $\sigma^2 \equiv \tfrac{1}{2}\, \sigma_{ab}\sigma^{ab}$, $\dot{u}^2 \equiv \dot{u}_{a}\dot{u}^{a}$.
Equation\z\eqref{Xdot1} shows how the exotic non-Newtonian nature of the effective 
Horndeski fluid is encoded in $\dot{X}$ and, therefore, in the fluid total pressure.
The first constitutive relation\z\eqref{constitutive} 
is automatically satisfied.

Substituting Eq.\z\eqref{Xdot} into Eq.\z\eqref{Xdot1} yields the system
\begin{align}
    F_{1}\,H&=A\,,\\
    &\nn\\
    F_{1}\,I + F_{2}\,H&=B\,,\\
    &\nn\\
    (F_{2}\,I-C)\,\Theta^2 - D\,\sigma^2 - E\,\dot{u}^2&=\,0\,.
\end{align}
The first two equations give unenlightening expressions of $F_{1}=A/H$ and
$F_{2}=(BH-AI)/H^2$, while the third one is a non-linear second
order differential equation for 
$\phi$. In general, the latter
(which is not derived from an action principle) is  
 incompatible with the field equation for $\phi$ (which is also of
second order) and cannot be imposed. For the same reason, we do not 
take into account the case $\dot{X}=0$, which corresponds to requiring
$\nabla_{a}\phi\nabla_{b}\phi\nabla^{a}\nabla^{b}\phi=0$.
The only way to implement self-consistently the requirement of a Newtonian effective fluid is to restrict the theory to $G_{3X}=0$. Then, $ G_{3}=G_{3}(\phi)$ 
and ${\cal L }_{3}=-G_{3}(\phi) \, \Box\phi$ can be 
absorbed into $G_{2}$ integrating by 
parts, effectively leading to $-G_{3}(\phi) \, \Box\phi=2 X\,G_{3\phi}$
plus a total divergence in the action. 
Then, the $\phi$-equation of motion in terms of kinematic 
quantities yields
\begin{align}
  F_1(\phi,X)=&\frac{\sqrt{2X} \left[ 2 G_2 G_{4\phi} + X G_{4\phi} \left( 6 
G_{4\phi\phi} - G_{2X} \right) \right]}{3 G_{4\phi }^2 + G_{4} \left( G_{2X } + 2 X 
G_{2XX } \right) }\nn\\
  &-\frac{\sqrt{2X} \, G_{4} \left( G_{2\phi} -2 X G_{2\phi X} \right)}{3 G_{4\phi 
}^2 
+ G_{4} \left( G_{2X } + 2 X G_{2XX } \right)}\,,\\
&\nn\\
F_2(\phi,X)=&- \frac{2 X \left( 3 G_{4\phi}^2 + G_{4} G_{2X} \right)}{3 
G_{4\phi}^2 + G_{4} \left( G_{2X} + 2 X G_{2XX } \right)}\,.
\end{align}
Only in this case Eq.\z\eqref{Xdot} is 
not an extra equation but coincides with the equation of 
motion of 
$\phi$. 

To recap, requiring that the effective fluid be linear in 
the gradient of its 4-velocity selects only two possible classes of 
Horndeski gravity. Either 
\begin{align}\label{class1}
  {\cal L}=G_{4}(\phi)R+G_{2}(\phi,X)-G_{4\phi}\ln{X}\,\Box\phi\,,
\end{align}
which corresponds to an effective fluid with
\begin{align}
  \rho=&\frac{1}{2G_4}(2XG_{2X}-G_{2}-2XG_{4\phi\phi}\ln{X})\,,\label{rho1}\\
  &\nn\\
  \bar{P}=&\frac{1}{2G_4}\left[ G_2+2XG_{4\phi\phi}\left( 2-\ln{X} \right) \right]\,,\label{P1}\\
  &\nn\\
  \eta=&-\frac{1}{2}\sqrt{2X}\,\frac{G_{4\phi}}{G_4}\,,\qquad 
  \zeta=\frac{2}{3}\sqrt{2X}\,\frac{G_{4\phi}}{G_4}\,,\\
  &\nn\\
  \xi=&\,0\,,\qquad q^{a}=0\,,
\end{align}
or else 
\begin{align}\label{class2}
  {\cal L}=G_{4}(\phi)R+G_{2}(\phi,X)\,,
\end{align}
which is instead characterized by
\begin{align}
  \rho=&\frac{1}{2G_4}\left( 2XG_{2X}-G_{2} \right),\label{rho2}\\
  &\nn\\
  \bar{P}=&- \frac{G_{4\phi}^2 \left( G_{2} -2 X G_{2X} \right)}{2 G_{4} \left[3 
G_{4\phi}^2 + G_{4} \left( G_{2X} + 2 X G_{2XX} \right) \right]}\nn\\
  &+ \frac{\left( G_{2} + 4 X G_{4\phi\phi} \right) \left( G_{2X} + 2 X G_{2XX} 
\right)}{2 \left[3 
G_{4\phi}^2 + G_{4} \left( G_{2X} + 2 X G_{2XX} \right) \right]}\nn\\
  &+ \frac{G_{4\phi} \left( G_{2\phi} -2 X G_{2\phi X} \right)}{3 G_{4\phi}^2 + 
G_{4} \left( G_{2X} + 2 X G_{2XX} \right) }\,,\label{P2}\\
&\nn\\
  \eta=&-\frac{1}{2}\sqrt{2X}\,\frac{G_{4\phi}}{G_4}\,,\qquad \xi=-\sqrt{2X} \, 
\, \frac{G_{4\phi}}{G_4}\,,\\
&\nn\\
  \zeta=&-\frac{\sqrt{2X} \, G_{4\phi} \left[ 3 G_{4\phi}^2 + G_{4} \left( G_{2X} -4 X 
G_{2XX} \right) \right]}{3 G_{4} \left[ 3 G_{4\phi}^2 + G_{4} \left( G_{2X} + 2 X 
G_{2XX} \right) \right]}\,,\label{zeta2}\\
&\nn\\
  q^{a}=&-\sqrt{2X} \, \frac{G_{4\phi}}{G_4} \, \dot{u}^{a}\label{q2}\,.
\end{align}
Both models reduce to GR if $G_{4}= 1$. They are two 
disconnected classes of Horndeski gravity in the sense that they are 
closed, and cannot change into each other, under disformal 
transformations.

In any situation different from these two cases, Horndeski gravity can be recast as an effective fluid characterized by the linear constitutive equations
\begin{align}
    \rho=&\,\bar{\rho}-\xi\Theta\,,\qquad  \pi_{\mu\nu}=-2\eta\,\sigma_{\mu\nu}\,,\qquad
    q^{\mu}=\xi\,\dot{u}^{\mu}
\end{align}
and by the non-Newtonian constitutive equation for the pressure which, using Eqs.\z\eqref{Xdot1}--\eqref{II} in Eq.\z\eqref{P}, can be parametrized as
    \begin{eqnarray}
        P&=&\dfrac{\bar{P}_{1}-\zeta_{1}\,\Theta-\zeta_{2}\,\Theta^2-\zeta_{3}\,\sigma^2-\zeta_{4}\,\dot{u}^2}{\bar{P}_{2}-\zeta_{5}\,\Theta \vphantom{A^{A^A}}}\,.
    \end{eqnarray}
For small velocity gradients $\nabla_{a}u_{b} $ the above constitutive equation reduces, to first order, the effective viable Horndeski fluid to one with Newtonian behaviour. 

Some additional comments are useful to conclude this part.

Only the first Horndeski class can admit a non-dynamical scalar field, {\it i.e.}, {an} extended cuscuton model\z\cite{Iyonaga:2018vnu} (corresponding to the subclass $G_3=G_{4\phi}\ln{X}$ {which implies} $ f_{3}=0$ in Ref.\z\cite{Miranda:2022brj}).

In the second Horndeski class, the denominators of Eqs.\z\eqref{P2} and\z\eqref{zeta2} vanish for theories with Lagrangian density 
\begin{align}\label{nondy}
    {\cal L}=G_{4} 
R+f_1(\phi)+f_2(\phi)\sqrt{2X}-\frac{3G_{4\phi}^2}{G_{4}} \, X\,,
\end{align}
which automatically excludes a non-dynamical scalar 
field, {\em i.e.}, the extended cuscuton model (corresponding to the subclass $G_3=0$, which implies $ f_{3}=-2f_{4\phi}$ in Ref.\z\cite{Miranda:2022brj})
and, in particular, pathological $\omega=-3/2$ Brans--Dicke gravity which corresponds 
to $f_1=f_2=0$ and $G_4=\phi$, as well as cuscuton gravity\z\cite{Afshordi:2006ad}. In this case, we cannot use the scalar field equation of motion to rewrite $\dot{X}$ (or $\Box{\phi}$) in terms of kinematic quantities because the corresponding multiplicative factor in Eq.\z\eqref{eom} vanishes identically. Therefore, one cannot write Eq.\z\eqref{Xdot}. 

The full action of the extended cuscuton model can be obtained by requiring the coefficients\z\eqref{HH} and\z\eqref{II} to vanish.
Moreover, the condition $G_{3X}+X\,G_{3XX}=0$ corresponds to the special case in which the viscous contribution to the pressure is a finite sum of terms at most quadratic in the 4-velocity gradient.

For the second linear Horndeski class it is possible to find the relation between the bulk viscosity and the energy density transport coefficient $\xi$
\begin{align}
    \zeta=\,\frac{\xi}{3}\left[\frac{3 G_{4\phi}^2 + G_{4} \left(G_{2X} -4 X  G_{2XX} \right) }{  3 G_{4\phi}^2 + G_{4}\left( G_{2X} + 2 X G_{2XX} \right) }\right]\,.
\end{align}
``First generation'' scalar-tensor gravity corresponds to $\zeta=\xi/3$ and $G_{4}=\phi$.

\subsection{General Horndeski gravity {\em in vacuo}}

We now briefly extend the previous  analysis to general Horndeski gravity, 
including $G_{4}(\phi,X)$ and $G_{5}(\phi, X)$. The 
classes of theories selected by imposing the Newtonian nature of the effective fluid are again given by Eqs.\z\eqref{class1} 
and\z\eqref{class2}. The linearity in $\nabla_a u_b$ implies 
$G_5=G_{4X}=0$ and then the above discussion holds. All we have 
to do is considering the energy density and total pressure 
for general Horndeski gravity and impose that all the non-linear terms vanish. The 
energy density is given by 
\begin{align}
  G_4\,\rho &=  \frac{2}{3} \, X \sqrt{2X} \, \Theta \, \sigma ^{2}{} 
\left( G_{5X} + X G_{5XX} \right) - \frac{2}{27} \, X \sqrt{2X} \, 
\Theta^3 \left( G_{5X} + X G_{5XX} \right)\nn\\
&\nn\\
&- \frac{2}{3} \, X \sqrt{2X} \, \sigma _{a}{}^{c} \sigma ^{ab} \sigma _{bc}
\left( G_{5X} + X G_{5XX} \right) \nn\\
&\nn\\
  & + X \sqrt{2X} \, R_{acbd} u^{a} u^{b}  \sigma ^{cd} G_{5X}+ 
\frac{2}{3} \, X 
\Theta^2 \left( G_{4X} + 2 X G_{4XX} - G_{5\phi} - X G_{5\phi X} 
\right)\nn\\
&\nn\\
  &+ 2 X \sigma^{2}{} \left( G_{5\phi} + X G_{5\phi X}- G_{4X} -2 
\sqrt{2X} \, \eta\, G_{5X} -2 X G_{4XX} \right) \nn\\
&\nn\\
  &- \frac{1}{2} \, \left( G_{2} -2 X G_{2X} + 2 X G_{3\phi} \right) + 
\rho \left[ X \left( 2 G_{4X} - G_{5\phi} \right) - \frac{X}{3} 
\, \sqrt{2X} \, G_{5X}\,\Theta\right] \nn\\
&\nn\\
&+\sqrt{2X} \, \Theta \left( G_{4\phi} + 2 X G_{4\phi X}- X G_{3X} \right)\,.
\end{align}
The first term in the second line of the above equation is multiplied by $R_{acbd} u^{a} u^{b}\sigma 
^{cd}=\dot{u}^{a} \dot{u}^{b} \, \sigma _{ab} - \tfrac{2}{3}\, \Theta \sigma 
^{2}{} - u^{a} \sigma ^{bc} \nabla _{a}\sigma _{bc} + \sigma _{ab} \nabla 
^{b}\dot{u}^{a} + u^{a} \sigma ^{bc} \nabla_{c}\sigma _{ab}$. Since the latter represents a non-linear contribution in $\nabla_a u_b$ that cannot be cancelled by any other term, it is immediate to see that $G_{5X}$ must vanish. Then, one can set 
$G_5=0$ because $G_5(\phi)$ can be absorbed, upon integration by parts, in the other functions $G_2, 
G_3$, and $G_4$ according to 
\begin{eqnarray}
G_{2}  \to  G_{2}-2X^2G_{5\phi\phi\phi} \,,\,\quad G_{3} \to  G_{3} -3XG_{5\phi\phi} \,,\,\quad G_4 \to  G_{4}-XG_{5\phi} \,.
\end{eqnarray}
The result is the Lagrangian density
\begin{eqnarray}
{\cal L}_{5} &=& G_5G_{ab}\nabla^{a}\nabla^{b}\phi\simeq 
-G_{5\phi}G_{ab}\nabla^{a}\phi\nabla^{b}\phi  -G_{5\phi}R_{ab}\nabla^{a}\phi\nabla^{b}\phi
-XG_{5\phi}R \nonumber\\
&\nn\\
&=& G_{5\phi}(\nabla_{a} 
\Box\phi-\nabla_{b}\nabla_{a}\nabla^{b}\phi) 
\nabla^{a}\phi-XG_{5\phi}R \nonumber\\
&\nn\\
& \simeq & - G_{5\phi\phi}(-2X\Box\phi-\nabla_{a}\nabla_{b}\phi 
\nabla^{a}\phi\nabla^{b}\phi)  -G_{5\phi}
\left[\Box\phi^2-(\nabla_{a} 
\nabla_{b}\phi)^2\right]-XG_{5\phi}R \nn\\
&\nn\\
& \simeq & 3X\,G_{5\phi\phi}\,\Box\phi-2X^2G_{5\phi\phi\phi}  -G_{5\phi}
\left[\Box\phi^2-(\nabla_{a} 
\nabla_{b}\phi)^2\right]-XG_{5\phi}R\,,
\end{eqnarray}
where $\simeq$ denotes equality up to a total divergence. The effective energy density then becomes
\begin{align}
    \rho  = & \frac{2 X \Theta ^2 \left( G_{4X} + 2 X G_{4XX} \right)}{3 
\left( G_{4} -2 X G_{4X} \right)} - \frac{2 X \sigma^{2}{} \left(G_{4X} + 
2 X G_{4XX} \right)}{G_{4} -2 X G_{4X} }\nn\\
&\nn\\
    & + \frac{\sqrt{2X} \, \Theta  \left( - X G_{3X} + G_{4\phi} + 2 X 
G_{4\phi X} \right)}{G_{4} -2 X G_{4X}}- \frac{G_{2} -2 X G_{2X} + 2 X G_{3\phi}}{2 G_{4} -4 X G_{4X}}
\end{align}
where, in order to suppress the quadratic terms in $\nabla_{a}u_b$ in the first line, it is necessary that 
\begin{equation}\label{G4}
    G_{4X}+2X G_{4XX}=0  \,,
\end{equation}
which implies that
\begin{equation}
G_4 \left( \phi, X \right) =G_{0} \left( \phi 
\right) + G_{1} \left( \phi \right) \sqrt{2X}\,.
\end{equation}
Using now Eq.\z\eqref{feq}, the perfect fluid contribution to the effective isotropic pressure becomes
\begin{align}
\bar{P} = & - \frac{2 X \Theta ^2 G_{4X}}{9 \left( G_{4} -2 X G_{4X} 
\right)}  + 
\frac{2 X \sigma^{2}{}  G_{4X}}{3 \left( G_{4} -2 X G_{4X} \right)} - 
\frac{4 \dot{u}^{2}{} X \left( G_{4X} + 2 X G_{4XX} \right)}{3 
\left(G_{4} -2 X G_{4X} \right)}\nn\\
&\nn\\
& + \frac{\dot{X} \left( -  X G_{3X} + G_{4\phi} + 2 X G_{4\phi 
X} \right) }{ \sqrt{2X}  \left( - G_{4}  + 2 X G_{4X}\right)}\nn\\
&\nn\\
&+ \Theta  \left[ \zeta  - \frac{2 
\dot{X} \left( G_{4X} + 2 X  G_{4XX} \right)}{3 \left( G_{4} -2 X G_{4X} 
\right)} + \frac{2 \sqrt{2X} \left( - G_{4\phi}  + 2 X G_{4\phi X} 
\right)}{3 \left( G_{4} -2 X G_{4X} \right)}\right]\nn\\
&\nn\\
     &+ \rho \, \frac{2 X  G_{4X}}{3 \left( G_{4} -2 X G_{4X} \right)}+ 
\frac{G_{2} -2 X 
G_{3\phi} + 4 X G_{4\phi\phi}}{2 G_{4} -4 X G_{4X}}
\end{align}
that, together with Eq.\z\eqref{G4}, requires $G_{4X}\left( \phi, X \right) =0$ to eliminate the quadratic terms in the first line. This is necessary 
because, even if we assume $\dot{X}$ quadratic in $\nabla_{a}u_{b}$ to cancel the quadratic terms, $\Theta\dot{X}$ in the second line reintroduces a cubic term. The theory then reduces to viable Horndeski, and the previous discussion remains valid.

\subsection{Horndeski gravity with matter}
 
Finally, let us include the matter in this picture. The Newtonian behaviour of the effective fluid requires again 
Eqs.~\eqref{class1} and\z\eqref{class2}, with the only
difference that the inviscid pressure\z\eqref{P2} acquires
the additional contribution 
\begin{align}
  \bar{P}\to \bar{P} - \frac{G_{4\phi}^2 \, T^\mathrm{(m)} }{G_{4} 
\left[ 3 G_{4\phi}^2 + G_{4} \left( G_{2X} + 2 X G_{2XX} \right) \right]}\,,
\end{align}
proportional to the trace of the matter energy-momentum tensor 
$T^\mathrm{(m)}=g^{ab}T^\mathrm{(m)}_{ab}$. The reason is that the presence of matter changes the field equations for $g_{ab}$ to
\begin{align}
    R_{ab}=&\,T_{ab}-\frac{T}{2} \,g_{ab}+\frac{1}{G_4}\left(T^\mathrm{(m)}_{ab}-\frac{T^\mathrm{(m)}}{2} \,g_{ab}\right)\,,\\
    &\nn\\
    R=&\,-T-\frac{T^\mathrm{(m)}}{G_{4}}\,,
\end{align}
turning the function $A(\phi,X)$ of Eq.\z\eqref{Xdot1} into
\begin{align}
    A\to A &+ \sqrt{2X}\,G_{4\phi}\,T^\mathrm{(m)}- \sqrt{2X}\left(T^\mathrm{(m)}+2T^\mathrm{(m)}_{ab}u^{a}u^{b}\right)X\,G_{3X}\,.
\end{align}

\section{Analogy with first-order general-relativistic viscous fluids}
\label{sec:3}

Before proceeding, let us recall the basic description of {\em real} dissipative fluids that will be applied to the Horndeski {\em effective} fluid later in this section.   
Dissipative fluids are out-of-equilibrium systems.
The most general stress-energy tensor describing an out-of-equilibrium system has the form\z\cite{Bemfica:2017wps,Bemfica:2019knx,Bemfica:2020zjp}
\begin{align}\label{genstress}
    T^{ab}= \left( \varepsilon+\mathcal{A} \right)\,u^{a}u^{b}+ \left( p +\mathcal{B} \right)\,h^{ab}+2q^{(a}u^{b)}+\pi^{ab}\,,
\end{align}
where $\mathcal{A}$ and $\mathcal{B}$ represent the out-of-equilibrium
corrections to $\eps$ and $p$, which are the equilibrium energy density and
pressure, respectively.  $\mathcal{A}$ and $\mathcal{B}$
vanish at equilibrium.
In the first-order formulation of viscous fluids\z\cite{Bemfica:2019knx,Kovtun:2019hdm,Hoult:2020eho},
all the quantities in Eq.\z\eqref{genstress} depend on the fluid 4-velocity $u^{a}$, 
the temperature $\T$, and the chemical potential $\mu$. In particular, the deviations
from equilibrium are parametrized by the gradients of $u^{a}$, $\T$, and $\mu$. Here we work in the Eckart (or particle) frame. In the effective fluid description,  we decompose the spacetime according to the effective fluid 4-velocity, therefore the fluid motion is described using the fluid's proper time.
If we consider vanishing chemical potential $\mu=0$, we can parametrize
the out-of-equilibrium quantities as
\begin{align}
\mathcal{A}&=\chi_1\dfrac{u^{a}\nabla_{a}\T}{\T}+\chi_2\,\Theta\,,\label{A}\\
&\nn\\
\mathcal{B}&=\chi_3\dfrac{u^{a}\nabla_{a}\T}{\T}+\chi_4\,\Theta\,,\label{B}\\  &\nn\\
q^{a}&=\lambda\left(\dfrac{h^{ab}\nabla_{b}\T}{\T}+\dot{u}^{a}\right)\,,\label{Q}\\
&\nn\\
\pi^{ab}&=-2\eta\,\sigma^{ab}\,,\label{PI}
\end{align}
where the transport coefficients $\chi_i$, $\lambda$, and $\eta$ depend on the temperature ${\cal T}$. Since the chemical potential vanishes identically, the equilibrium energy density and pressure depend only on the temperature, $\eps=\eps(\T)$ and $p=p(\T)$. We  parametrize $\lambda$ as $\lambda=-\K\,\T$, where $\K=\K(\T)$ is the thermal conductivity\z\cite{Eckart:1940te,Faraoni:2021lfc,Faraoni:2018qdr} of the effective fluid.

The second law of thermodynamics then yields\z\cite{Baumann:2022mni}
 \begin{align}
     \frac{dp}{d\T}=\dfrac{\eps+p}{\T}\,,\label{dp/dT}
 \end{align}
which can also be written as
\begin{align}
  \eps(\T)=-p(\T)+\T\,p'(\T)\,, \label{eps(T)}
\end{align}
where a prime denotes differentiation with respect to the temperature (see Appendix\z\ref{thermo}).

Let us apply now the thermodynamic fluid description to the Horndeski {\em effective} fluids classes\z\eqref{class1} and\z\eqref{class2} in the framework of first-order general-relativistic viscous fluids. This procedure allows one to discuss the ``temperature of gravity'' and its transport coefficients. 
Before discussing the individual Horndeski classes, let us make some considerations valid in both cases.

In general, $\T=\T(\phi,X)$ and 
\begin{align}
     \nabla_{a}\T=\left(\sqrt{2X}\,\T_{\phi}-\T_X\,\dot{X}\right)u_{a}-2X\,\T_X \,\dot{u} _{a}\,,
\end{align}
therefore Eqs.\z\eqref{A}--\eqref{Q} turn into
\begin{align}
    \mathcal{A}&=\chi_1 \, \dfrac{-\sqrt{2X}\,\T_{\phi}+\T_{X}\,\dot{X}}{\T}+\chi_2\,\Theta\,,\label{Ah}\\
    &\nn\\
    \mathcal{B}&=\chi_3 \, \dfrac{-\sqrt{2X}\,\T_{\phi}+\T_{X}\,\dot{X}}{T}+\chi_4\,\Theta\,,\label{Bh}\\
    &\nn\\
    q^{a}&=\lambda\left(-2X \, \frac{\T_{X}}{\T} +1\right)\dot{u}^{a}\,.\label{Qh}
\end{align}
In both classes we have 
\begin{align}
    \eta=-\frac{\sqrt{2X}}{2}\frac{G_{4\phi}}{G_4}\,,   \label{ETA}
\end{align}
and the shear viscosity $\eta(\T)$ depends on the temperature. Its  derivative with respect to $X$ is  
\begin{equation}
    \eta'(\T) \, \T_X=-\frac{1}{2\sqrt{2X}} \, \frac{G_{4\phi}}{G_4} \,,
\end{equation}
implying that $T_X$ is always non-vanishing unless $G_{4\phi} = 0$. We  rewrite this equation as
\begin{align}
    2X\,\T_X=\frac{\eta(\T)}{\eta'(\T)}\,,
\end{align}
where $2X\,\T_X$ still depends on the temperature. Eq.~(\ref{dp/dT}) implies
\begin{align}\label{dp}
    p_{\phi}=\frac{\T_{\phi}}{\T}\left( \eps+p \right)\,,\qquad p_{X}=\frac{\T_{X}}{\T}\left( \eps+p \right)\,.
\end{align}
If $p=-\eps$ identically, the above relations imply constant equilibrium pressure and energy density, $p=-\eps=-\Lambda$.

In both cases, we note a further constraint on the linear Horndeski classes: the functional form of $G_2$ is determined up to an unknown function of $F=F(\sqrt{2X}\,\tfrac{G_{4\phi}}{G_{4}})$ (see Eqs.\z\eqref{G2:class1} and\z\eqref{G2:class2}). While in the first class we obtain $\T=\alpha\,\sqrt{2X}\,\tfrac{G_{4\phi}}{G_{4}}$ and then  $F=F\left(\T\right)\equiv p(\T)$ (the equilibrium pressure), in the second one we can parameterize the dependence of $F$ using the shear viscosity in Eq.\z\eqref{ETA}, with the result that  $F(\T)=p(\T)-4\eta^2(\T)$. 

\subsection{Class I: \texorpdfstring{$G_3=G_{4\phi}\ln{X}$}{G3=G4philnX} }

Comparing Eqs.\z\eqref{PI},\z\eqref{Ah}--\eqref{Qh} with the quantities associated to the class\z\eqref{class1}, one obtains
\begin{align}
    \chi_1&=\chi_2=\chi_3=0\,,\qquad 3\chi_{4}=4\eta\,,\\
    &\nn\\
    \chi_4&=-\frac{2}{3}\,\sqrt{2X}\,\frac{G_{4\phi}}{G_4}\,, \qquad\eta=-\frac{1}{2}\,\sqrt{2X}\,\frac{G_{4\phi}}{G_4}\,,\\
    &\nn\\
    \eps&=\frac{1}{2G_4}, (2XG_{2X}-G_{2}-2X G_{4\phi\phi} \ln{X})\,,\\
    &\nn\\
    p&=\frac{1}{2G_4}\left[G_2+2XG_{4\phi\phi} \left( 2-\ln{X}\right)\right]\,,
\end{align}
and
\begin{align}
    \lambda\left(-2X \, \frac{\T_{X}}{\T} +1\right)=0\,.\label{q=0}
\end{align}
This equation is satisfied when the bracket is equal to zero or when $\lambda \equiv {\cal K}{\cal T} =0$. In the first case, the solution of the differential equation is
\begin{align}
    \T(\phi,X)=\sqrt{2X}\,C(\phi)
\end{align}
and, we can find the integrating function $C(\phi)$ by imposing $\eta=\eta(\T)$, 
\begin{align}
    \T(\phi,X)=\alpha\,\sqrt{2X}\,\frac{G_{4\phi}}{G_{4}}\,,\label{T1}
\end{align}
where $\alpha$ is a constant.\footnote{ In order for the effective temperature to be positive-definite, $G_4$ must be monotonic and  $\alpha\,G_{4\phi}>0$. If an effective Newton constant decreasing in time is assumed, then $G_{4\phi}>0$, $\alpha>0$, and $\eta<0$. Negative viscosity is characteristic of non-isolated systems, that exchange energy
with their surroundings.} Then the Horndeski function $G_{2}$ has the form
\begin{align}
    G_2 \left( \phi, X \right)= 2G_4 \,F\left(\sqrt{2X}\,\frac{G_{4\phi}}{G_{4}}\right)-2X\,G_{4\phi\phi}(2-\ln{X})\,,    \label{G2:class1}
\end{align}
where $F=F(\T)\equiv p(\T)$. The equilibrium energy density $\eps(\T)$ is automatically given by Eq.\z\eqref{eps(T)}.

The situation $G_{4\phi}=0$ corresponds to $\T=0$ and  $p(0)=-\eps(0)=-\Lambda$. 
The Horndeski Lagrangian density collapses into $R-2\Lambda$, where $\Lambda$ is the cosmological constant.

In particular, a linear relation between pressure and temperature 
\begin{align}
    p(\T)&=-\gamma+\beta\,\T=-\gamma+\alpha\,\beta\,\sqrt{2X}\,\frac{G_{4\phi}}{G_{4}}\,,
\end{align}
identifies the extended cuscuton model with  $f_1=-2\gamma\,G_{4}$, $f_2=2\,\alpha\,\beta\,G_{4\phi}$, and $f_3=0$ (following the notation of Ref.\z\cite{Miranda:2022brj}), where $\alpha$, $\beta$, and $\gamma$ are constants. The energy density turns into 
\begin{align}
     \eps(\T)&=\gamma+\beta\,\T=\gamma+\alpha\,\beta\,\sqrt{2X}\,\frac{G_{4\phi}}{G_{4}}\,.
\end{align}

Assuming a linear barotropic equation of state $p=w\,\eps$, where $w$ is the constant equation of state parameter, we obtain the energy density
\begin{equation}
    \eps(\T)=\gamma\,\T^{\frac{w+1}{w}}
\end{equation}
and
\begin{equation}
    G_2 = 2w\,\gamma\,G_4\left(\alpha\sqrt{2X}\,\frac{G_{4\phi}}{G_{4}}\right)^{\frac{w+1}{w}}-2X\,G_{4\phi\phi}(2-\ln{X})\,.
\end{equation}
If $(\T-2X\,\T_{X})\neq0$, Eq.\z\eqref{q=0} is satisfied only for $\lambda=0$. This is compatible only with constant pressure and energy density, $p=-\eps=-\Lambda$, which corresponds to the form of $G_2$
\begin{align}
    G_2 (\phi, X)  = -2\Lambda\,G_4-2X\,G_{4\phi\phi}(2-\ln{X})\,.
\end{align}
The Horndeski Lagrangian coincides with a particular extended cuscuton model given by $f_1=-2\Lambda\,G_{4}\,$, $f_2=0$, and $f_3=0$ (see again the notation of\z\cite{Miranda:2022brj}). Therefore, we have a non-dynamical imperfect fluid that mimics the cosmological constant and whose inviscid/equilibrium contributions satisfy the familiar linear barotropic equation $p=-\eps$. 

\subsection{Class II: \texorpdfstring{$G_3=0$}{G3=0}}

For the second class, we obtain the system
\begin{align}
    \chi_1&=0\,,\qquad \chi_2\chi_3+\chi_3\lambda+\chi_2\lambda=0\,,\qquad 3\chi_{4}=4\eta\,,\\
    &\nn\\
    \chi_2&=\sqrt{2X}\,\frac{G_{4\phi}}{G_{4}}\,,\qquad\chi_4=-\frac{2}{3}\sqrt{2X}\,\frac{G_{4\phi}}{G_{4}}\,,\\
    &\nn\\
    \chi_3&=-\frac{1}{\sqrt{2X}}\,\frac{G_{4\phi}}{G_{4}}\,\frac{\T}{\T_{X}}\,,\\
    &\nn\\
    \lambda&=-\sqrt{2X}\,\frac{G_{4\phi}}{G_{4}}\,\frac{\T}{\T-2X\,\T_{X}}\,,\\
    &\nn\\
    \eta&=-\frac{1}{2}\,\sqrt{2X}\,\frac{G_{4\phi}}{G_4}\,,\label{etaT}\\
    &\nn\\
    \eps&=\frac{2X\,G_{2X}-G_{2}}{2G_{4}}\,,\label{ee}\\
    &\nn\\
    p&=\frac{G_{2}+4X\,G_{4\phi\phi}}{2G_{4}}-\frac{G_{4\phi}}{G_{4}}\frac{\T_{\phi}}{T_{X}}\,.
\end{align}
In this case, Eq.\z\eqref{T1} cannot be a solution for the temperature, {\it i.e.}, $\T$ cannot be proportional to $\eta$, otherwise we would have $\T-2X\,\T_X=0$ and vanishing heat flux. 

Using Eq.\z\eqref{etaT}, we write the partial derivatives of $\T$ in terms of $\eta(\T)$,
\begin{align}
    \T_X&=\frac{1}{2X}\frac{\eta(\T)}{\eta'(\T)}\,,\\
    &\nn\\
    \T_\phi&=\frac{2}{\sqrt{2X}}\frac{\eta^2(\T)}{\eta'(\T)}-\frac{1}{2}\frac{\sqrt{2X}}{\eta'(\T)}\frac{G_{4\phi\phi}}{G_4}\,,
\end{align}
which yields
\begin{align}
    \frac{\T_\phi}{\T_X}=-2X\,\frac{G_{4\phi}}{G_4}+2X\,\frac{G_{4\phi\phi}}{G_{4\phi}}\,.\label{Tphi/Tx}
\end{align}
Then the pressure reads
\begin{align}\label{pp}
    p=\frac{G_{2}}{2G_{4}}+2X\,\frac{G_{4\phi}^2}{G_{4}^2}\,,
\end{align}
or  
\begin{align}
    G_2&=2G_4\,p(\T)-2X\,\frac{G_{4\phi}^2}{G_{4}}=2G_4\left[\,p(\T)-4\eta^2(\T)\right]\,.
\end{align}
Substituting this expression of $G_2$ in the equilibrium energy density\z\eqref{ee}, one finds  
\begin{align}
    \eps(\T)&=-p(\T)+2X\,\T_X\,p'(\T)-2X\frac{G_{4\phi}^2}{G_{4}^2}\nn\\
    &=-p(\T)+2X\,\T_X\,p'(\T)-4\eta^2(\T)\,.\label{epsT}
\end{align}
We can rewrite $\eps+p$ in the above equation using Eq.\z\eqref{dp/dT} to obtain
\begin{align}
    (\T-2X\,\T_X)\,p'=-4\eta^2
\end{align}
while, substituting $p'=(\eps+p)/\T$ and Eqs.\z\eqref{ee} and\z\eqref{pp} in Eq.\z\eqref{epsT} yields the temperature
\begin{align}
    \T=C(\phi)\,\exp{\left[\int^{X}_{1}\frac{4G_{4\phi}^2+G_4\,G_{2Z}}{2Z\left(2G_{4\phi}^2+G_4\,G_{2Z}\right)}\,dZ\right]}\,,
\end{align}
where $C(\phi)$ is a generic integration function of the scalar field\footnote{ $C(\phi)$ can be determined  {\it a posteriori} by imposing $\eta=\eta(\T)$.} and the integration is performed with respect to $Z$, an auxiliary variable of the kinetic scalar.
Finally, the correspondence $\lambda=-\K\T$ gives
\begin{align}
     (\T-2X\,\T_X)\,\K=-2\eta\,,\quad\mathrm{and}\quad p'=2\eta\,\K\,.
\end{align}

The relations\z\eqref{dp} and\z\eqref{Tphi/Tx} yield
\begin{align}
    \frac{p_\phi}{p_X}=\frac{\T_\phi}{\T_X} \,,
\end{align}
providing the system
\begin{align}
    G_2&=2G_4\,F(\eta)\,,\label{G2:class2}\\
    &\nn\\
    p(\eta)&=F(\eta)+4\,\eta^2\,,\label{p(eta)}\\
    &\nn\\
    \eps(\eta)&=-F(\eta)+\eta\,\frac{dF(\eta)}{d\eta}\label{eps(eta)}\,,
\end{align}
where, to avoid non-dynamical scalar fields, one must have $F(\eta)\neq\alpha+\beta\,\eta-3\eta^2$  with $\alpha$ and $\beta$ constant (see Eq.\z\eqref{nondy}). 

Imposing Eq.\z\eqref{dp/dT} on $\eps(\eta)$ and on $p(\eta)$ is equivalent to 
\begin{align}\label{F(eta)}
   4\eta^2+\eta\frac{dF}{d\eta}=\T\left(8\eta+\frac{dF}{d\eta}\right)\eta' \,,
\end{align}
which is rewritten as
\begin{align}\label{Q(T)}
   4\eta^2+\eta\,\frac{d\T}{d\eta}\,Q'=\T\left(8\eta+\frac{d\T}{d\eta}\,Q'\right)\eta'
\end{align}
where $Q(\T)=F(\eta ( \T ))$. 
If $\eta$ is a monotonic function of the temperature, Eq.\z\eqref{Q(T)} turns into
\begin{equation}
    4\eta^2+\frac{\eta}{\eta'} \, Q'=\T\left(8\eta\,\eta'+Q'\right)\,.\label{Q'(T)}
\end{equation}

If we consider a linear barotropic equation of state $p=w\,\eps$, Eq.\z\eqref{eps(T)} gives  
\begin{align}
    \eps&=\gamma\,T^{\frac{w+1}{w}}\,,
\end{align}
where the effective temperature is positive defined and $\gamma>0$ and $w$ are constant, while Eqs.\z\eqref{p(eta)} and\z\eqref{eps(eta)} yield the  differential equation
\begin{align}
    F(\eta)+4 \eta ^2=w \left(\eta\,\frac{dF(\eta)}{d\eta}-F(\eta)\right)\,.
\end{align}
In terms of the new variable $x=\eta^2$, this equation reads
\begin{equation}
    2wx\,\frac{dF(x)}{dx}-(1+w)F(x)-4x=0\,,
\end{equation}
which is recognized as an Euler--Cauchy differential equation with solutions
\begin{align}
    F_1(x)&=\frac{4 x}{w-1}+c_1\, x^{\frac{w+1}{2 w}}\,, &w\neq1\,,\\
    &\nn\\
    F_2(x)&=2 x \log (x)+c_2\, x\,, &w=1\,,
\end{align}
where $c_{1,2}$ are integration constants. In correspondence of $F_{1,2}$ we have the following expressions for the temperature
\begin{align}
    T_1=& \gamma^{-\frac{w}{w+1}}\left(\frac{4 x}{w-1}+c_1\,\frac{x^{\frac{w+1}{2 w}}}{w}\right)^{\frac{w}{w+1}}\,,\\
    &\nn\\
    T_2=& \left[\gamma^{-1}\,x\,(4+c_2+2 \ln{x})\right]^{1/2}\,,
\end{align}
respectively.  Let us consider now the power-law form
\begin{align}
    \T= \left( \alpha \, \eta \right)^\beta\,,
\end{align}
where $\alpha<0$ and $\beta\neq0,1,$ are constant.\footnote{In this case, $\alpha<0$, $\eta<0$, and $G_4$ is strictly monotonic.} Equation\z\eqref{F(eta)} then yields 
\begin{align}
    F(x)&=c_1-\frac{2 (\beta -2) x}{\beta -1}\,,
\end{align}
where $c_1$ is an integration constant, and $x=\eta^2$. The equilibrium energy density and pressure become, respectively,
\begin{align}
\eps&=-c_1-\frac{2 (\beta -2)}{\beta -1}\,x\,,\\
    &\nn\\
    p&=c_1+\frac{2 \beta}{\beta -1}\,x\,. 
\end{align}    
From the above expressions, for $\eps$ to be positive, it must be $c_1\leq0$ and $\beta <1 $ or $ \beta \ge 2$.

Our last example consists of the linear relation
\begin{align}
    \T=\alpha\,\eta+\beta \,,
\end{align}
which implies
\begin{align}
    F(\eta)=&-\frac{4 \alpha}{3 \beta }\,\eta ^3-4 \eta ^2+c_1\,,\\
    &\nn\\
    p(\eta)=&-\frac{4 \alpha  \eta ^3}{3 \beta }+c_1\,,\\
    &\nn\\
    \eps(\eta)=&-\frac{8 \alpha  \eta ^3}{3 \beta }-4 \eta ^2-c_1\,,
\end{align}
where $c_1$ is constant. In this case,  $\T$ and $G_{2}$ become constant as $\eta \to 0$ and the energy density is positive-definite for $ c_1<0$. 

\begin{table*}[!hb]
\centering
\caption{Overview of viable Horndeski effective fluids.}
\resizebox{\textwidth}{!}{\begin{NiceTabular}{|llrllllcclll|}
\hline
\multicolumn{12}{|c|}{\textbf{\textsc{Viable Horndeski as an effective fluid}\xrowht{20pt}}} \\ \hline
\multicolumn{12}{|c|}{Assumption of timelike scalar field gradient:\xrowht{15pt}} \\ 
\multicolumn{12}{|l|}{} \\
\multicolumn{12}{|c|}{$\displaystyle u^{a} =\frac{\nabla ^{a} \phi }{\sqrt{2X}}\,,\quad\quad X=-\frac{1}{2} \nabla _{a} \phi \nabla ^{a} \phi  >0\,.$} \\
\multicolumn{12}{|l|}{} \\ \hline
\multicolumn{12}{|c|}{Stress-energy tensor of a dissipative fluid:\xrowht{15pt}} \\ 
\multicolumn{12}{|l|}{} \\
\multicolumn{12}{|c|}{\begin{tabular}[c]{@{}c@{}}$ T^{ab} =\rho \, u^{a} u^{b} +P\, h^{ab} +2\, q^{( a} u^{b)} +\pi ^{ab} \,,$\\ \\  $\quad\displaystyle \rho =$ total energy density$\,$, $\displaystyle \quad P=$ total isotropic pressure$\,$, $\displaystyle \quad q^{a} =$ heat flux density$\,$, $\displaystyle \quad \pi ^{ab} =$ anisotropic stress$\,$.$\quad$\end{tabular}} \\
\multicolumn{12}{|l|}{} \\ \hline
\multicolumn{12}{|c|}{Stress-energy tensor of viable Horndeski effective fluid (no assumptions):\xrowht{15pt}} \\ 
\multicolumn{12}{|l|}{} \\
\multicolumn{12}{|c|}{\begin{tabular}[c]{@{}c@{}}$\displaystyle T^{ab} =(\bar{\rho } -\xi \, \Theta ) u^{a} u^{b} +\left(\bar{P} +\frac{4}{3} \ \eta \, \Theta +\xi \, \frac{\dot{X}}{2X}\right) h^{ab} +2\ \xi \, \dot{u}^{( a} u^{b)} -2\, \eta \, \sigma ^{ab}$,\\ \\ $\displaystyle \eta =-\frac{1}{2} \, \sqrt{2X} \, \frac{G_{4\phi }}{G_{4}}\,,$$\displaystyle \quad\quad \xi =\sqrt{2X} \, \frac{( G_{4\phi } -X\, G_{3X})}{G_{4}}\,,$\\ \\ where $\displaystyle \bar{\rho }$, $\displaystyle \bar{P}$, $\displaystyle \xi $, $\displaystyle \eta $ are functions of $\displaystyle \phi $ and $\displaystyle X$ and an overbar denotes \av non-viscous\cv contributions; \\ $\displaystyle \eta $ is unambiguously interpreted as shear viscosity coefficient.\end{tabular}} \\
\multicolumn{12}{|l|}{} \\ \cline{3-10}
$\quad\quad\quad\quad\quad\quad\quad\quad$ &
  \multicolumn{1}{l|}{$\quad\quad\quad\quad\quad\quad\quad$} &
  \multicolumn{5}{r}{$\displaystyle\rho =$} &
  \multicolumn{1}{l|}{$\displaystyle \bar{\rho } -\xi \, \Theta$ \xrowht{20pt}} &
  \multicolumn{2}{c|}{$\quad$Linear in $\displaystyle \nabla _{a} u_{b}$$\quad$} &
   &
   \\ \cline{3-10}
 &
  \multicolumn{1}{l|}{} &
  \multicolumn{5}{r}{$\displaystyle P=$} &
  \multicolumn{1}{l|}{$\displaystyle \bar{P} +\frac{4}{3} \, \eta \, \Theta +\xi \, \dfrac{\dot{X}}{2X}\quad\quad$\xrowht{30pt}} &
  \multicolumn{2}{c|}{Not explicit} &
   &
   \\ \cline{3-10}
 &
  \multicolumn{1}{l|}{} &
  \multicolumn{5}{r}{$\displaystyle q^a=$} &
  \multicolumn{1}{l|}{$\displaystyle \xi \, \dot{u}^{a}$\xrowht{20pt}} &
  \multicolumn{2}{c|}{Linear in $\displaystyle \nabla _{a} u_{b}$} &
   &
   \\ \cline{3-10}
 &
  \multicolumn{1}{l|}{} &
  \multicolumn{5}{r}{$\quad\displaystyle \pi^{ab}=$} &
  \multicolumn{1}{l|}{$\displaystyle -2\,\eta \, \sigma ^{ab}$\xrowht{20pt}} &
  \multicolumn{2}{c|}{Linear in $\displaystyle \nabla _{a} u_{b}$} &
   &
   \\ \cline{3-10}
\multicolumn{12}{|l|}{} \\ \hline
\multicolumn{12}{|l|}{} \\
\multicolumn{12}{|c|}{\begin{tabular}[c]{@{}c@{}}$\quad\quad\quad$To understand the nature of the Horndeski effective fluid ({\it i.e.}, to obtain the constitutive equations of the fluid)$\quad\quad\quad$\\ we must be able to write the total pressure in terms of kinematic quantities.\end{tabular}} \\
\multicolumn{12}{|l|}{} \\ \hline
\multicolumn{7}{|c|}{$\boldsymbol{\xi =0}$ \xrowht{25pt}} &
  \multicolumn{5}{c|}{$\displaystyle \boldsymbol{\xi \neq 0}$ (only  dynamical scalar fields)\xrowht{25pt}} \\ \hline
\multicolumn{7}{|c|}{} &
  \multicolumn{3}{c}{$\displaystyle P=\frac{\bar{P}_{1} -\zeta _{1} \, \Theta -\zeta _{2} \, \Theta ^{2} -\zeta _{3} \, \sigma ^{2} -\zeta _{4} \, \dot{u}^{2}}{\bar{P}_{2} -\zeta _{5} \, \Theta \vphantom{A^{A^A}}}\,,$\xrowht{48pt}} &
  \multicolumn{2}{c|}{\begin{tabular}[c]{@{}c@{}}Non-Newtonian (in general)$\quad$\\$\bar{P}_i$ and $\zeta_i$ functions of $\phi$ and $X$$.$$\quad$\end{tabular}} \\ \cline{8-12} 
\multicolumn{7}{|c|}{\multirow{-5}{*}{\begin{tabular}[c]{@{}c@{}}\textbf{Linear Horndeski - Class I:}\xrowht{13pt}\\\\ $G_3(\phi,X)=G_{4\phi}\ln{X}$\\\\Landau frame and Eckart frame\\  coincide.\end{tabular}}} &
  \multicolumn{5}{c|}{} \\
\multicolumn{7}{|c|}{$\,$\xrowht{27pt}} &
  \multicolumn{5}{c|}{\multirow{-2}{*}{\textbf{Linear Horndeski - Class II: $G_3(\phi,X)=0$}}} \\ \hline
\multicolumn{7}{|c|}{\begin{tabular}[c]{@{}c@{}}{$\begin{aligned}[t] &\\ \rho  & =\bar{\rho }\,,\\ P & =\bar{P} -\zeta \, \Theta \,,\quad\quad \left(\zeta =-\frac{4}{3} \eta\right) \\ q^{a} & =0\,,\\\pi ^{ab} & =-2\, \eta \, \sigma ^{ab}\,.\\& \end{aligned}$}\end{tabular}} &
  \multicolumn{5}{c|}{\begin{tabular}[c]{@{}c@{}}{$\begin{aligned}[t]&\\ \rho  & =\bar{\rho } -2\, \eta \, \Theta\,, \\ P & =\bar{P}^{*} -\zeta ^{*} \, \Theta \, =\bar{P} +\frac{4}{3} \eta \, \Theta +\eta \, \frac{\dot{X}}{2X}\,,\quad\quad\left(\bar{P}^*=\frac{\bar{P}_1}{\bar{P}_2}\,,\quad\zeta^*=\frac{\zeta_1}{\bar{P}_{2}}\right)\\ q^{a} & =2\, \eta \, \dot{u}^{a}\,,\\ \pi ^{ab} & =-2\, \eta \, \sigma ^{ab}\,.\\& \end{aligned}$}\end{tabular}} \\ \hline
\multicolumn{12}{|l|}{} \\
\multicolumn{12}{|c|}{\begin{tabular}[c]{@{}c@{}}The coefficients $\displaystyle \zeta $ and $\displaystyle \zeta ^{*}$ are the bulk viscosity coefficients for the class I and II, respectively.\\ The two classes above are the only linear effective fluid classes in the general Horndeski theory.\end{tabular}} \\
\multicolumn{12}{|l|}{} \\ \hline
\end{NiceTabular}}
\label{tab:1}
\end{table*}

\begin{table*}[!ht]
\caption{Analogy between linear Horndeski effective fluids and first-order viscous fluids.}
\centering
\resizebox{\textwidth}{!}{\begin{NiceTabular}{|cccccc|}
\hline
\multicolumn{6}{|c|}{\textbf{\textsc{Linear Horndeski effective fluids as first-order viscous fluids in Eckart frame}}} \\ \hline
\multicolumn{6}{|c|}{First-order thermodynamics of viscous fluids\xrowht{15pt}} \\ \hline
\multicolumn{6}{|c|}{$\quad$Stress-energy tensor of a dissipative fluid in Eckart's theory (vanishing chemical potential  $\mu=0$):$\quad$\xrowht{15pt}} \\
\multicolumn{6}{|c|}{\begin{tabular}[c]{@{}c@{}}$\,$\xrowht{5pt}\\ $\displaystyle T^{ab} =\varepsilon \, u^{a} u^{b} +( p-\zeta \, \Theta ) \, h^{ab} +2\, q^{( a} u^{b)} -2\eta \, \sigma ^{ab}$\,,\\ \\ $\displaystyle \varepsilon =$ equilibrium energy density$\,$,$\,\,\,\quad\displaystyle p=$ equilibrium pressure$\,$,$\,\,\,\quad\zeta =$ bulk viscosity$\,$,$\,\,\,\quad\displaystyle \eta =$ shear viscosity$\,$, \\ \\ $\displaystyle q^{a} =-\mathcal{K}\left( h^{ab} \nabla _{b}\mathcal{T} +\mathcal{T}\dot{u}^{a}\right)\,,$\\ \\ where $\K=\K(\T)$ is the thermal conductivity and $\T$ is the fluid temperature.\\ $\,$\xrowht{5pt}\end{tabular}} \\ \hline
\multicolumn{6}{|c|}{General first-order stress-energy tensor of viscous fluids ($\kern+0.01em\displaystyle \mu =0\kern+0.01em $):\xrowht{15pt}} \\
\multicolumn{6}{|c|}{\begin{tabular}[c]{@{}c@{}}\xrowht{4pt}\\ $T^{ab} =( \varepsilon +\mathcal{A}) \, u^{a} u^{b} +( p+\mathcal{B}) \, h^{ab} +2\, q^{( a} u^{b)} +\pi ^{ab}\,,$\\ $\,$\xrowht{6pt}\end{tabular}} \\
\multicolumn{6}{|c|}{${\quad\mathcal{A} =\displaystyle \chi _{1} \, \frac{u^{a} \nabla _{a}\mathcal{T}}{\mathcal{T}} +\chi _{2} \, \Theta\,,\quad\quad\mathcal{B} =\displaystyle \chi _{3} \, \frac{u^{a} \nabla _{a}\mathcal{T}}{\mathcal{T}} +\chi _{4} \, \Theta\,,\quad\quad q^{a}=\displaystyle \lambda \,  \left( \, \frac{h^{ab} \nabla _{b}\mathcal{T}}{\mathcal{T}} +\dot{u}^{a}\right),\quad\quad\pi ^{ab} =\displaystyle -2\eta \, \sigma ^{ab}\,,\quad}$} \\
\multicolumn{6}{|c|}{\begin{tabular}[c]{@{}c@{}}$\,$\xrowht{6pt}\\ $\displaystyle p'(\T)=\dfrac{\varepsilon(\T)+p(\T)}{\T}\,,$\\ \\ where all transport coefficients $\displaystyle \chi _{i}$, $\displaystyle \eta $, and $\displaystyle \lambda =-\mathcal{KT}$ are functions of the temperature $\displaystyle \mathcal{T}$.\\ Eckart's theory is obtained for $\displaystyle \chi _{1} =\chi _{2} =\chi _{3} =0\,$.\\ $\,$\xrowht{5pt}\end{tabular}} \\ \hhline{|======|}
\multicolumn{6}{|c|}{} \\ \cline{2-3} \cline{5-5}
\multicolumn{1}{|c|}{} &
  \multicolumn{1}{c|}{$\quad$Horndeski effective fluid\xrowht{15pt}$\quad$} &
  \multicolumn{1}{c|}{$\quad$First-order fluids\xrowht{15pt}$\quad$} &
  \multicolumn{1}{c|}{} &
  \multicolumn{1}{c|}{$\displaystyle\Rightarrow\quad\T=\T(\phi,X)\,,\quad\T_{X}\neq0$\xrowht{15pt}} &
   \\
\multicolumn{1}{|c|}{\multirow{-2}{*}{$\quad\quad\quad$}} &
  \multicolumn{1}{c|}{$\displaystyle\eta=-\frac{1}{2}\,\sqrt{2X}\,\frac{G_{4\phi}}{G_4}$\xrowht{25pt}} &
  \multicolumn{1}{c|}{$\displaystyle\eta=\eta(\T)$\xrowht{25pt}} &
  \multicolumn{1}{c|}{$\quad\enspace$} &
  \multicolumn{1}{c|}{$\quad\displaystyle\nabla_a\T=\left(\sqrt{2X}\,\T_\phi-\dot{X}\,\T_{X}\right)u_a-2X\,\T_{X}\,\dot{u}_a\quad$} &
   \\ \cline{2-3} \cline{5-5}
\multicolumn{6}{|c|}{\xrowht{10pt}} \\
\multicolumn{6}{|c|}{{$\displaystyle\mathcal{A}=\chi _{1} \, \frac{-\sqrt{2X} \, \mathcal{T}_{\phi } +\dot{X} \, \mathcal{T}_{X}}{\mathcal{T}} +\chi _{2} \, \Theta\,,\quad\quad \mathcal{B}=\chi _{3} \, \frac{-\sqrt{2X} \, \mathcal{T}_{\phi } +\dot{X} \, \mathcal{T}_{X}}{\mathcal{T}} +\chi _{4} \, \Theta\,,\quad\quad q^{a}=\lambda \,  \left( -2X\, \frac{\mathcal{T}_{X}}{\mathcal{T}} +1\right)\dot{u}^{a}\,.$}} \\
\multicolumn{6}{|c|}{\xrowht{10pt}} \\ \hline
\multicolumn{3}{|c|}{\textbf{Linear Horndeski - Class I \xrowht{15pt}}} &
  \multicolumn{3}{c|}{\textbf{Linear Horndeski - Class II \xrowht{15pt}}} \\ \hline
\multicolumn{3}{|c|}{$\chi _{1} =\chi _{2} =\chi _{3} =0,\quad 3\chi _{4} =4\eta$ \xrowht{25pt}} &
  \multicolumn{3}{c|}{\begin{tabular}[c]{@{}c@{}}$\displaystyle \chi _{1} =0,\quad \chi _{2} =-2\eta ,\quad 3\chi _{4} =4\eta\,,$\\ $\displaystyle \chi _{2} \chi _{3} +\chi _{2} \lambda +\chi _{3} \lambda =0$\end{tabular}} \\ \hline
\multicolumn{3}{|c|}{\begin{tabular}[c]{@{}c@{}}From the condition $\displaystyle q^{a} =0$ we obtain \ $\displaystyle \mathcal{T} \varpropto \eta\,$:\\ \\ $\displaystyle \mathcal{T} =\alpha \,\displaystyle \sqrt{2X} \, \frac{G_{4\phi }}{G_{4}}\,,$\\ \\ where $\displaystyle \alpha $ is constant.\\\\Then, $\displaystyle \varepsilon $ and $\displaystyle p$ automatically satisfy $\displaystyle \varepsilon +p=\mathcal{T} \, p'$, and\\ \\ $\displaystyle G_{2} =2G_{4} \, p(\mathcal{T}) -2X\, G_{4\phi \phi }( 2-\ln X)\,,$\\\\ or, equivalently,\\ \\ $\,\,\,\displaystyle G_{2} =2G_{4} \, p({\eta}) -2X\, G_{4\phi \phi }( 2-\ln X)\,.$\end{tabular}} &
  \multicolumn{3}{c|}{\begin{tabular}[c]{@{}c@{}}$\,$\xrowht{4pt}\\It is not possible obtain the explicit form of $\displaystyle \mathcal{T}$\\ $\quad$ without making some assumption. In full generality: $\quad$\\\\
  $\displaystyle\T=C(\phi)\,\exp{\left[\int^{X}_{1}\frac{4G_{4\phi}^2+G_4\,G_{2Z}}{2Z\left(2G_{4\phi}^2+G_4\,G_{2Z}\right)}\,dZ\right]}\,,$\\\\
  and\\
  ${G_{2}}{=2\,G_{4}\left[\,p(\eta)-4\eta^2\right]}\,, $\xrowht{25pt}\\ $\varepsilon(\eta){=-p( \eta ) +\eta\left(\dfrac{dp}{d\eta}-4\eta\right)}\,,$\\ \\ $\displaystyle \eta\left(\dfrac{dp}{d\eta}-4\eta\right) =\dfrac{dp}{d\eta}\eta'\,\mathcal{T}\,.$\\ $\,$\end{tabular}} \\ \hline
\multicolumn{6}{|c|}{} \\
\multicolumn{6}{|c|}{\begin{tabular}[c]{@{}c@{}}The above analogy provides a constraint on the linear Horndeski classes:\\ $\displaystyle G_{2}$ is specified up to a function of the shear viscosity $p(\eta)$, the equilibrium pressure of the effective fluid.\end{tabular}} \\
\multicolumn{6}{|c|}{} \\ \hline
\end{NiceTabular}}
\label{tab:2}
\end{table*}

\section{Conclusion}\label{sec12}

It is hard to overemphasize the importance of Horndeski gravity in the current research on gravitational theory. However, due to their complexity, physical insight in this class of theories still lags behind formal developments, and the effective fluid approach offers a new physical view.
We provide a physical interpretation of the Horndeski effective fluid as a relativistic non-Newtonian fluid.
The linearity of the effective fluid stress-energy tensor in the 4-velocity gradient selects two disconnected classes of viable Horndeski,
which guarantees that gravitational waves
propagate at the speed of light (see Table\z\ref{tab:1}). In any other Horndeski 
theory, it is impossible to recast the energy-momentum tensor in the
form~\eqref{stress} with viscosity coefficients linear in $\nabla_{a}u_{b}$ as in the usual constitutive equations. The interpretation of such theories is that their effective fluid equivalent is an exotic non-Newtonian fluid. This physical characterization of Horndeski theories of gravity based on the associated effective fluid has escaped attention thus far. Moreover, other alternative theories of gravity may admit a similar classification. This correspondence opens up a new research direction, which consists of looking for specific Horndeski theories that implement particular non-Newtonian rheologies of real fluids in their associated effective fluid. In conjunction with observations, such a search could potentially restrict the wide spectrum of Horndeski theories.

In general, the analogy between fluids and gravity can represent a pragmatic tool to understand current theoretical and observational problems. Finally, interest in the classes corresponding to Newtonian fluids is motivated by the possibility of representing the limit {of a more complete theory} for small 4-velocity gradients, similar to the way that classical theories are low-energy limits of effective field theories. In particular, in the context of late-time cosmology where the scalar field can be thought of as a low-energy field, this interpretation offers an additional tool for discriminating classes of Horndeski gravity.
We have applied our result to first-order viscous fluids and discussed their thermodynamics, which provides a further constraint on the linear Horndeski classes (see Table\z\ref{tab:2}).  In our analysis, theories in which the extra scalar degree of freedom (in addition to the usual two spin-2 modes of GR) does not propagate still occupy a special place in the classification of Horndeski theories based on the Newtonian versus non-Newtonian character of the effective equivalent fluid. In this light, they perhaps deserve reconsideration.

\backmatter

\bmhead{Acknowledgments}

M.\z M., D.\z V., and S.\z C. acknowledge the support of Istituto Nazionale di Fisica Nucleare (INFN) {\it iniziative specifiche} TEONGRAV, QGSKY, and MOONLIGHT2. D.\z V. also acknowledges the FCT project with ref. number PTDC/FIS-AST/0054/2021. M.\z M. is grateful to Bishop's University for the hospitality. V.\z F. is supported by the Natural Sciences \& Engineering Research Council of Canada (grant 2016-03803).
\begin{itemize}
\item Conflict of interest/Competing interests: The authors declare no conflict of interest
\item Ethics approval: Not applicable
\item Consent to participate: Not applicable
\item Consent for publication: Not applicable
\item Availability of data and materials: No Data associated with the manuscript
\item Code availability: Not applicable
\item Authors' contributions: All the Authors contributed equally to this work
\end{itemize}

\noindent

\bigskip

\begin{appendices}

\section{Thermodynamic equation for energy density and pressure at equilibrium}
\label{thermo}

From the second law of thermodynamics for particles in equilibrium at $\displaystyle \mu =0$, it follows that
\begin{equation}
    \varepsilon (\mathcal{T}) +p(\mathcal{T}) =\mathcal{T} p'(\mathcal{T})\,,
\end{equation}
where $\displaystyle S=S( V,\mathcal{T})$ is the entropy of the system (a function of its volume and temperature),  $\displaystyle U=\varepsilon V$ is the total equilibrium energy, and a prime denotes differentiation with respect to the temperature. Then  we have 
\begin{align}
dS&=\frac{1}{\mathcal{T}}( \, dU+p\, dV\, )\nn\\
&=\frac{1}{\mathcal{T}}[ \, d( \varepsilon \kern+0.01em V\, ) \ +p\, dV\, \, ]         \nn\\
&=\frac{1}{\mathcal{T}}[ V\, d\varepsilon +( \varepsilon +p) \kern+0.01em dV\,]\,,
\end{align}

\begin{equation}
\frac{\partial S}{\partial V} =\frac{1}{\mathcal{T}}( \varepsilon +p) \,,\quad \quad \,\quad \frac{\partial S}{\partial \mathcal{T}} =\frac{V}{\mathcal{T}} \, \varepsilon'   \,.
\end{equation}

The entropy differential $\displaystyle dS$ is exact and, therefore, closed,  $\tfrac{\partial^2 S}{\partial \mathcal{T} \partial V} =\tfrac{\partial^2 S}{\partial V\partial \mathcal{T} } \,,$ giving
\begin{align}
\frac{\partial }{\partial \mathcal{T}}\left[ \, \frac{1}{\mathcal{T}}( \varepsilon +p) \, \right] &  =\frac{\partial }{\partial V} \left( \, \frac{V}{\mathcal{T}} \, \varepsilon '\kern+0.01em \right)  \,, \nn\\
-\frac{1}{\mathcal{T}^{2}}(  \varepsilon +p) +\frac{1}{\mathcal{T}}( \varepsilon '+p') &  =\frac{1}{\mathcal{T}}\,\varepsilon'  \,, \nn
\end{align}
and finally
\begin{align}
p' & \displaystyle =\frac{1}{\mathcal{T}}\,( \varepsilon +p) \,.
\end{align}




\end{appendices}



\begin{thebibliography}{9}

\bibitem{Capozziello:2002rd}
S.~Capozziello,
``Curvature quintessence,''
Int. J. Mod. Phys. D \textbf{11}, 483-492 (2002)
doi:10.1142/S0218271802002025.

\bibitem{Faraoni:2010pgm}
V.~Faraoni and S.~Capozziello,
``Beyond Einstein Gravity: A Survey of Gravitational Theories for Cosmology and Astrophysics,''
Springer, New York, 2011,
doi:10.1007/978-94-007-0165-6.

\bibitem{Horndeski:1974wa}
G.~W.~Horndeski,
``Second-order scalar-tensor field equations in a four-dimensional space,''
Int. J. Theor. Phys. \textbf{10}, 363-384 (1974)
doi:10.1007/BF01807638.

\bibitem{Kobayashi:2019hrl}
T.~Kobayashi,
``Horndeski theory and beyond: a review,''
Rept. Prog. Phys. \textbf{82}, no.8, 086901 (2019)
doi:10.1088/1361-6633/ab2429.

\bibitem{Ostrogradsky:1850fid}
M.~Ostrogradsky,
``M\'emoires sur les \'equations diff\'erentielles, relatives au probl\`eme des isop\'erim\`etres,''
Mem. Acad. St. Petersbourg \textbf{6}, no.4, 385-517 (1850).

\bibitem{Woodard:2015zca} 
R.~P.~Woodard, 
``Ostrogradsky's theorem on Hamiltonian instability,'' 
Scholarpedia \textbf{10}, no.8, 32243 (2015) 
doi:10.4249/scholarpedia.32243.

\bibitem{Wald:1984rg}
R.~M.~Wald,
``General Relativity,''
Chicago Univ. Press, Chicago, 1984,
doi:10.7208/chicago/9780226870373.001.0001.

\bibitem{LIGOScientific:2017vwq} 
B.~P.~Abbott \textit{et al.} [LIGO Scientific and Virgo], 
``GW170817: Observation of Gravitational Waves from a Binary Neutron Star Inspiral,'' 
Phys. Rev. Lett. \textbf{119}, no.16, 161101 (2017) 
doi:10.1103/PhysRevLett.119.161101.

\bibitem{LIGOScientific:2017zic} 
B.~P.~Abbott \textit{et al.} [LIGO Scientific, Virgo, Fermi-GBM and INTEGRAL], 
``Gravitational Waves and Gamma-rays from a Binary Neutron Star Merger: GW170817 and 
Astrophys. J. Lett. \textbf{848}, no.2, L13 (2017) 
doi:10.3847/2041-8213/aa920c.

\bibitem{Creminelli:2017sry} 
P.~Creminelli and F.~Vernizzi,
``Dark Energy after GW170817 and GRB170817A,''
Phys. Rev. Lett. \textbf{119}, no.25, 251302 (2017)
doi:10.1103/PhysRevLett.119.251302.

\bibitem{Baker:2017hug}
T.~Baker, E.~Bellini, P.~G.~Ferreira, M.~Lagos, J.~Noller and I.~Sawicki,
``Strong constraints on cosmological gravity from GW170817 and GRB 170817A,''
Phys. Rev. Lett. \textbf{119}, no.25, 251301 (2017)
doi:10.1103/PhysRevLett.119.251301.

\bibitem{Bettoni:2016mij} 
D.~Bettoni, J.~M.~Ezquiaga, K.~Hinterbichler and M.~Zumalac\'arregui,
``Speed of Gravitational Waves and the Fate of Scalar-Tensor Gravity,'' 
Phys. Rev. D \textbf{95}, no.8, 084029 (2017) 
doi:10.1103/PhysRevD.95.084029.

\bibitem{Andreou:2019ikc}
N.~Andreou, N.~Franchini, G.~Ventagli and T.~P.~Sotiriou,
``Spontaneous scalarization in generalised scalar-tensor theory,''
Phys. Rev. D \textbf{101}, no.10, 109903(E) (2020)
doi:10.1103/PhysRevD.99.124022.

\bibitem{Nucamendi:2019uen}
U.~Nucamendi, R.~De Arcia, T.~Gonzalez, F.~A.~Horta-Rangel and I.~Quiros,
``Equivalence between Horndeski and beyond Horndeski theories and imperfect fluids,''
Phys. Rev. D \textbf{102}, no.8, 084054 (2020)
doi:10.1103/PhysRevD.102.084054.

\bibitem{Faraoni:2018qdr} 
V.~Faraoni and J.~C\^ot\'e, 
``Imperfect fluid description of modified gravities,'' 
Phys. Rev. D \textbf{98} (2018) 
no.~8, 084019 doi:10.1103/PhysRevD.98.084019.


\bibitem{Jacobson:1995ab}
T.~Jacobson,
``Thermodynamics of space-time: The Einstein equation of state,''
Phys. Rev. Lett. \textbf{75}, 1260-1263 (1995)
doi:10.1103/PhysRevLett.75.1260.

\bibitem{Eling:2006aw}
C.~Eling, R.~Guedens and T.~Jacobson,
``Non-equilibrium thermodynamics of spacetime,''
Phys. Rev. Lett. \textbf{96}, 121301 (2006)
doi:10.1103/PhysRevLett.96.121301.

\bibitem{Chirco:2010sw}
G.~Chirco, C.~Eling and S.~Liberati,
``Reversible and Irreversible Spacetime Thermodynamics for General Brans-Dicke Theories,''
Phys. Rev. D \textbf{83}, 024032 (2011)
doi:10.1103/PhysRevD.83.024032.

\bibitem{Faraoni:2021lfc}
V.~Faraoni and A.~Giusti,
``Thermodynamics of scalar-tensor gravity,''
Phys. Rev. D \textbf{103}, no.12, L121501 (2021)
doi:10.1103/PhysRevD.103.L121501.

\bibitem{Giusti:2021sku}
A.~Giusti, S.~Zentarra, L.~Heisenberg and V.~Faraoni,
``First-order thermodynamics of Horndeski gravity,''
Phys. Rev. D \textbf{105}, no.12, 124011 (2022)
doi:10.1103/PhysRevD.105.124011.

\bibitem{Giardino:2022sdv}
S.~Giardino, V.~Faraoni and A.~Giusti,
``First-order thermodynamics of scalar-tensor cosmology,''
JCAP \textbf{04}, no.04, 053 (2022)
doi:10.1088/1475-7516/2022/04/053.


\bibitem{Eckart:1940te}
C.~Eckart,
``The Thermodynamics of irreversible processes. 3.. Relativistic theory of the simple fluid,''
Phys. Rev. \textbf{58}, 919-924 (1940)
doi:10.1103/PhysRev.58.919.

\bibitem{Iyonaga:2018vnu}
A.~Iyonaga, K.~Takahashi and T.~Kobayashi,
``Extended Cuscuton: Formulation,''
JCAP \textbf{12}, 002 (2018)
doi:10.1088/1475-7516/2018/12/002.

\bibitem{Miranda:2022brj}
M.~Miranda, D.~Vernieri, S.~Capozziello and V.~Faraoni,
``Generalized McVittie geometry in Horndeski gravity with matter,''
Phys. Rev. D \textbf{105}, no.12, 124024 (2022)
doi:10.1103/PhysRevD.105.124024.

\bibitem{Afshordi:2006ad}
N.~Afshordi, D.~J.~H.~Chung and G.~Geshnizjani,
``Cuscuton: A Causal Field Theory with an Infinite Speed of Sound,''
Phys. Rev. D \textbf{75}, 083513 (2007)
doi:10.1103/PhysRevD.75.083513.

\bibitem{Bemfica:2017wps}
F.~S.~Bemfica, M.~M.~Disconzi and J.~Noronha,
``Causality and existence of solutions of relativistic viscous fluid dynamics with gravity,''
Phys. Rev. D \textbf{98}, no.10, 104064 (2018)
doi:10.1103/PhysRevD.98.104064.

\bibitem{Bemfica:2019knx}
F.~S.~Bemfica, M.~M.~Disconzi and J.~Noronha,
``Nonlinear Causality of General First-Order Relativistic Viscous Hydrodynamics,''
Phys. Rev. D \textbf{100}, no.10, 104020 (2019)
[erratum: Phys. Rev. D \textbf{105}, no.6, 069902 (2022)]
doi:10.1103/PhysRevD.100.104020.

\bibitem{Bemfica:2020zjp}
F.~S.~Bemfica, M.~M.~Disconzi and J.~Noronha,
``First-Order General-Relativistic Viscous Fluid Dynamics,''
Phys. Rev. X \textbf{12}, no.2, 021044 (2022)
doi:10.1103/PhysRevX.12.021044.

\bibitem{Kovtun:2019hdm}
P.~Kovtun,
``First-order relativistic hydrodynamics is stable,''
JHEP \textbf{10}, 034 (2019)
doi:10.1007/JHEP10(2019)034.

\bibitem{Hoult:2020eho}
R.~E.~Hoult and P.~Kovtun,
``Stable and causal relativistic Navier-Stokes equations,''
JHEP \textbf{06}, 067 (2020)
doi:10.1007/JHEP06(2020)067.

\bibitem{Baumann:2022mni}
D.~Baumann,
``Cosmology,''
Cambridge University Press, Cambridge, 2022,
doi:10.1017/9781108937092.


\end{thebibliography}


\end{document}